# Investigation of two-color magneto-optical trap with cesium $6S_{1/2}$-$6P_{3/2}$-$7S_{1/2}$ ladder-type system

Jie Wang[1, 2], Guang Yang[1, 2], Baodong Yang[1, 2], Jun He[1, 2, 3], and Junmin Wang[1, 2, 3,*]

1. State Key Laboratory of Quantum Optics and Quantum Optics Devices (Shanxi University),
2. Institute of Opto-Electronics, Shanxi University,
3. Collaborative Innovation Center of Extreme Optics (Shanxi University),
No.92 Wu Cheng Road, Tai Yuan 030006, Shan Xi Province, People's Republic of China

E-mail: wwjjmm@sxu.edu.cn (Junmin Wang)

**Abstract:** A novel cesium (Cs) two-color magneto-optical trap (TC-MOT), which partially employs the optical radiation forces due to photon scattering of the $6P_{3/2}$ (F'=5) - $7S_{1/2}$ (F"=4) excited-state transition in the Cs $6S_{1/2}$ - $6P_{3/2}$ - $7S_{1/2}$ (852 + 1470 nm) ladder-type system, has been proposed and experimentally investigated. One of the three pairs of 852 nm cooling/trapping beams (CTBs) in a conventional Cs MOT is replaced with a pair of the 1470 nm CTBs (type-I) or with one 852 nm CTB plus another counter-propagating 1470 nm CTB (type-II). Both the type-I and type-II Cs TC-MOTs can cool and trap atoms on both the red- and blue-detuning sides of the two-photon resonance. The Cs TC-MOT demonstrated in this work may have applications in the background-free detection of cooled and trapped atoms, and the photon-pair sources compatible with the ensemble-based quantum memory and the long-distance quantum communication via optical fiber.

## 1. Introduction

Laser cooling and trapping of neutral atoms played an important role, and caused a profound impact in many fields such as precision measurements, optical atomic clock, quantum degenerate gases, quantum information processing and so on **[1-2]**. Up to now, most laser cooling schemes used the optical radiation forces due to photon scattering from the single-photon transition between atomic ground states and excited states. This approach has been extremely successful, leading to a range of techniques including the Doppler cooling **[3]**, the polarization gradient cooling **[4]**, and the velocity-selective coherent population trapping [5]. However, there are a few theoretical and experimental investigations of two-photon laser cooling in a ladder-type atomic system. Furthermore, these studies mostly focused on the cooling of alkaline-earth-metal atoms as a second stage using the narrow $^1S_0$-$^3P_1$ inter-combination transition after initial precooling with the strong $^1S_0$-$^1P_1$ dipole transition [6-8]. Most recently, a two-color magneto-optical trap (TC-MOT) based on cesium (Cs) $6S_{1/2}$ - $6P_{3/2}$ - $8S_{1/2}$ (852 + 795 nm) ladder-type system, which partially uses the optical radiation forces due to photon scattering between two excited states ($6P_{3/2}$ and $8S_{1/2}$), has been demonstrated experimentally [9, 10]. The TC-MOT can cool and trap atoms on both the red- and blue-detuning sides of the two-photon resonance. This approach has been applied to a background-free detection of trapped atoms from the related transitions driven by no laser beam with a help of narrow-bandwidth high-contrast interference filters in our previous work [10]. Also, this approach has application in assisting cooling of certain atomic or molecular species which require lasers at inconvenient wavelengths. For instance, a laser cooling technique to cool hydrogen or anti-hydrogen atoms using cooling transition between excited states was proposed [11].

---
[*] Author to whom any correspondence should be addressed.

More potential and distinctive applications of the TC-MOT may be the photon-pair source compatible with the ensemble-based quantum memory and the long-distance quantum communication via optical fiber [12, 13]. For many years, the spontaneous parametric down conversion in a nonlinear crystal has become the standard method for generating entangled photon pairs [14, 15]. This kind of photon-pair source has limitations for certain applications due to the broad line-width (~ GHz) and the short coherence time (sub-picosecond). In 2005, Balić et al [16] reported a cold rubidium (Rb) ensemble based four-wave mixing to generate paired photons with coherence time of ~ 50 ns and line-width of ~ 9 MHz. They used a four-level system with two hyperfine ground states to generate the Stokes and anti-Stokes photons at 780 and 795 nm, respectively. Then they improve the results by using a two-dimensional MOT with a higher optical depth [17, 18]. However, previously described photon sources do not offer much flexibility with the photon wavelength. In particular, it is not possible to directly create photons at 1.5 μm. In 2006, Chaneliere et al [13] demonstrated 1529 and 780 nm entangled photon pairs by using of a cascade transition of a ladder-type system in a cold Rb ensemble. 1529 nm photons are ideal for the long-distance quantum communication due to low loss in silica fiber, while 780 nm photons are naturally suitable for storing quamtun information into and retrieving it from a long-lived Rb quantum memory. Inspired by this idea, the Cs TC-MOT based on $6S_{1/2}$ - $6P_{3/2}$ - $7S_{1/2}$ (852 + 1470 nm) ladder-type system may be an alternative approach of the photon-pair source compatible with the long-lived Cs quantum memory and the long-distance quantum communication via optical fiber. The 852 and 1470 nm laser beams in the Cs TC-MOT will not only serve as the cooling/trapping beams (CTBs), but also serve as the pump beams of four-wave mixing for the paired-photon generation. In detail, there are two possible ways: (1) One possible way is that the CW pair-production experiment is directly operated in type-II TC-MOT. The 852 and 1470 nm CTBs counter-propagate with a small angle (for example, ~1°, it almost does not affect the TC-MOT operation), serve as both the CTBs in the cooling and trapping process and the pump beams in the four-wave mixing, while the 852 nm CTBs in x-y plane with low power take part in the cooling and trapping process. In previous paired-photon source with a cold ensemble [16-18], the photon generation window of 500 μs is followed by a cooling and trapping period of 4.5 ms. The duty cycle of 10% surely reduces the pair generation rate and limits application. Maybe our paired-photon source can overcome this limitation. However, it is not clear that pair-production experiments could easily be performed in the complicated polarization configuration and the multiple beams of a MOT in any case;

(2) Another possible way is that we firstly cool and trap atoms in type-II TC-MOT, and then turn off the CTBs in x-y plane, while the CTBs along z axis remain for four-wave mixing. At least this arrangement can save physical space (do not need additional pump beams any more) and laser powers.

The primary motivation of this work is not only for the 852 and 1470 nm photon-pair source compatible with the Cs ensemble-based quantum memory and the long-distance quantum communication via optical fiber, but also for the understanding of cooling and trapping mechanism from the multi-photon transitions. Fig. 1(a) shows the decay channels from Cs $8S_{1/2}$ and $7S_{1/2}$ states, besides the different decay rate ($\gamma = 2\pi \times 3.30$ MHz for $7S_{1/2}$, and $\gamma = 2\pi \times 1.52$ MHz for $8S_{1/2}$), the significant difference is that: compared with the Cs $6S_{1/2}$ - $6P_{3/2}$ - $8S_{1/2}$ TC-MOT **[9, 10]**, the Cs $6S_{1/2}$ - $6P_{3/2}$ - $7S_{1/2}$ TC-MOT is significantly simpler due to less decay channels. Consequently, it is profitable for the understanding of cooling and trapping mechanism in this simple and different ladder-type system. To be specific, the behaviors are a bit different between them in type-I TC-MOT, see Fig. 2(a) and the discussion.

In addition, the main virtue of less decay channels is that the trapped atom's number is much more linearly dependent on the fluorescence from the TC-MOT to a certain degree. More strictly, the relationship between the atom number and the fluorescence is not perfectly linear due to different optical power, different detuning, non-cyclic transition, and so on. However, the fluorescence indicates a large number of atoms are trapped in the MOT. Consequently, we directly measure the fluorescence to diagnose whether the TC-MOT operates and determine the range of two-photon detuning for the TC-MOT operation.

In this article, a novel Cs TC-MOT based on Cs $6S_{1/2}$ - $6P_{3/2}$ - $7S_{1/2}$ (852 + 1470 nm) ladder-type system is proposed and experimentally investigated, which partially employs the optical radiation forces due to photon scattering of the $6P_{3/2}$ (F'=5) - $7S_{1/2}$ (F"=4) excited-state transition. One of the three pairs of 852 nm CTBs in a conventional Cs MOT is replaced with a pair of the 1470 nm CTBs (type-I) or with one 852 nm CTB plus another counter-propagating 1470 nm CTB (type-II). Both the type-I and type-II Cs TC-MOTs can cool and trap atoms on both the red- and blue-detuning sides of the two-photon resonance. We measured and analyzed qualitatively the dependence of peak fluorescence (cold atom number) on the two-photon detuning, the intensity of CTBs and the different combination of CTBs along the z direction (the axis of the anti-Helmholtz coils of the TC-MOT). These results provide optimized experimental parameters to trap atoms and pave the first step towards application of type-II Cs TC-MOT for 852 and 1470 nm entangled photon pair generation.

The rest of this paper is organized as follows. In Sec. 2 we briefly introduce the experimental setup of type-I and type-II Cs TC-MOTs and discuss the cooling and trapping effects. In Sec. 3 we develop a frequency offset locking system to control the detuning of two lasers. In Sec. 4 we measured and analyzed qualitatively the dependence of peak fluorescence on the two-photon detuning, the intensity of CTBs in type-I and type-II TC-MOT. Lastly, we draw our concluding remarks in Sec. 5.

**2. Experimental apparatus and principle**

In this section, we firstly introduce the relevant energy-level transitions and the experimental setup of type-I and type-II Cs TC-MOTs, and then we discuss the cooling and trapping effects.

*2.1. Experimental apparatus*

Fig. 1 shows the relevant energy-level transitions and the schematic diagram of two type of laser beam configuration for the TC-MOT. Fig. 1(b) shows an energy-level diagram of relevant transitions. The 852 nm CTBs interacted with |g>–|e> transition ($\omega_1$) has a detuning of $\Delta_1$. The 1470 nm CTBs interacted with |e>–|e'> excited-state transition ($\omega_2$) have a detuning of $\Delta_2$ and the two-photon detuning is $\delta_2$. The 852 nm repumping beams resonant with $6S_{1/2}$ (F=3) - $6P_{3/2}$ (F'=4) transition. The excited state line-widths for |e> and |e'> states are $\Gamma/2\pi = 5.2$ MHz and $\gamma/2\pi = 3.3$ MHz [19], respectively. We consider two types of CTB configuration for the Cs TC-MOT: type-I as shown in Fig. 1(c), the CTBs in the x-y plane comprises the two pairs of conter-propagating 852 nm beams (~ 14 mm, $1/e^2$ diameter), the CTBs along the z axis (the axis of the anti-Helmholtz coils of the TC-MOT) is a pair of conter-propagating 1470 nm beams (~ 10 mm, $1/e^2$ diameter); type-II as shown in Fig. 1(d), the CTBs in the x-y plane is the same as type-I, the CTBs along the z axis comprises one 852 nm CTB (~ 14 mm, $1/e^2$ diameter) and one counter-propagating 1470 nm CTB (~ 10 mm, $1/e^2$ diameter). The 852 nm repumping beams (not shown in Fig. 1) with ~ 16 mm $1/e^2$ diameter are sent along the ±y axis. In our experiment, the gradient of the quadrupole magnetic field generated by a pair of anti-Helmholtz coils with current $I$ is 1.0 mT/cm (10 Gauss/cm) along the z direction. The pressure of Cs atomic vapor inside the stainless steel vacuum chamber is $1.06\times10^{-6}$ Pa ($8\times10^{-9}$ Torr).

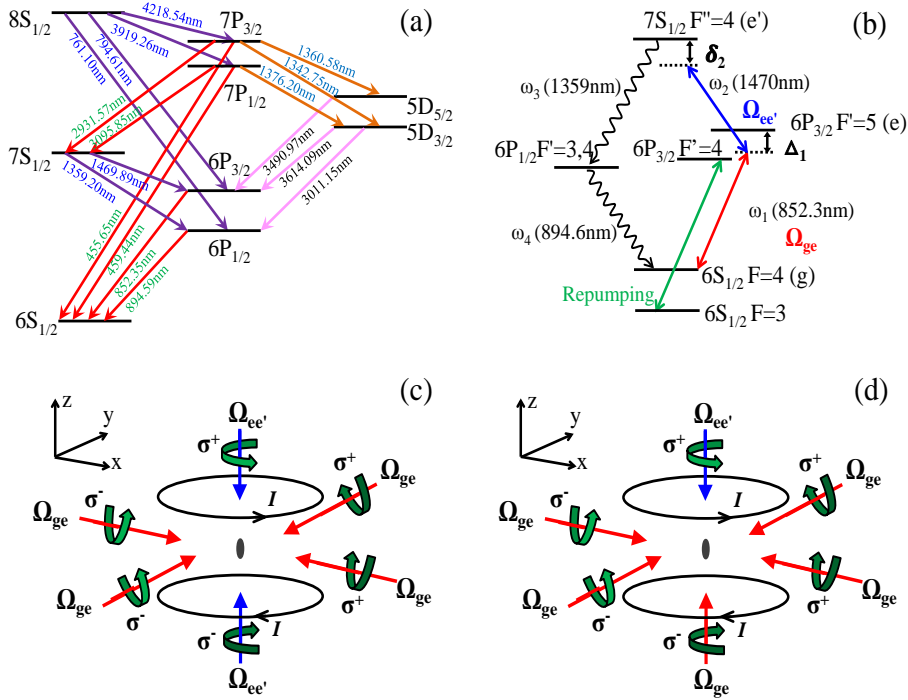

**FIG. 1.** (Color online) Relevant energy-level transitions and schematic diagram of two type of laser beam configuration for the TC-MOT. (a) Relevant energy-level transitions of Cs atoms. There are very less decay channels from $7S_{1/2}$ state than from $8S_{1/2}$ state, which is profitable to the analysis of cooling mechanism. (b) The transitions interacting with the 852 and 1470 nm CTBs and repumping beams. (c) type-I TC-MOT, two 1470 nm CTBs counter-propagate along the z axis, (d) type-II TC-MOT, one 1470 nm CTB and one 852 nm CTB counter-propagate along the z axis, $\sigma^\pm$ are specified with respect to the positive x, y and z axis, and $I$ is the electric current of the anti-Helmholtz coils.

## 2.2. Principle

The cooling and trapping process in type-I TC-MOT arises from two effects [9]: one is the velocity-dependent scattering force, associated with 2-photon or 3-photon scattering process; another is the position-dependent restoring force, which is essential for trapping. Here the restoring force was found when the helicities of the |e>–|e'> CTBs are opposite to those for the conventional MOT. The restoring force has the correct sign for both positive and negative δ$_2$ when Δ$_1$ < 0 **[9]**. We mainly introduce the velocity-dependent scattering force as following.

*2.2.1. Two-photon scattering process*

In low intensity regime, the dominant radiation pressure along the z axis is due to 2-photon scattering, where the first photon is absorbed from the in the x-y plane laser beams and the second photon is absorbed from the CTBs along the z axis. The scattering forces along the z axis can be written as $\mathbf{f}_z^{(2)} = \hbar k_{ee'} \sum_{\mathbf{i},\mathbf{j}} R_{\mathbf{i},\mathbf{j}}^{(2)} \hat{\mathbf{j}}$, where $\hat{\mathbf{i}} \in \{\hat{\mathbf{x}}, -\hat{\mathbf{x}}, \hat{\mathbf{y}}, -\hat{\mathbf{y}}\}$ is one of the four directions of the |g>–|e> CTBs, and $\hat{\mathbf{j}} \in \{\hat{\mathbf{z}}, -\hat{\mathbf{z}}\}$ is one of the two directions of the |e>–|e'> CTBs. For a Cs atom moving with a velocity $\mathbf{v}$, the 2-photon scattering rate in the low intensity limit can be written as:

$$R_{\mathbf{i},\mathbf{j}}^{(2)} = \frac{\gamma |\Omega_{ge} \Omega_{ee'}|^2}{16 |(\tilde{\Delta}_1 - k_{ge} \hat{\mathbf{i}} \cdot \mathbf{v})(\tilde{\delta}_2 - k_{ge} \hat{\mathbf{i}} \cdot \mathbf{v} - k_{ee'} \hat{\mathbf{j}} \cdot \mathbf{v})|^2} . \qquad (1)$$

Here $\Omega_{ge}$ and $\Omega_{ge}$ are the Rabi frequencies of the laser induced couplings per beam; $k_{ge}$ and $k_{ee'}$ are the wave numbers of the |g>–|e> and |e>–|e'> CTBs; $\tilde{\Delta}_1 = \Delta_1 + i\Gamma/2$ and $\tilde{\delta}_2 = \delta_2 + i\gamma/2$; Taylor-expanding Eq. (1) around $v_z = \hat{\mathbf{z}} \cdot \mathbf{v} = 0$ gives $f_z^{(2)} \approx -\alpha^{(2)} v_z$, with the 2-photon damping coefficient $\alpha^{(2)} > 0$ for the negative 2-photon detuning $\delta_2 < 0$. This is similar to the Doppler cooling process in the conventional MOT, where the Doppler effect enhances the absorption cross section for the |e>–|e'> CTBs opposing the velocity $\mathbf{v}$.

*2.2.2. Three-photon scattering process*

At moderate and high intensity of the |g>–|e> CTBs, the cooling also works, which is opposed to the 2-photon scattering cooling. This should be attributed to 3-photon and higher order scattering process. In the 3-photon process, the 2-photon absorption is followed by a stimulated |e'>–|e> emission. These multi-photon processes can lead to efficient cooling along the z axis in a manner similar to the "Doppleron" cooling **[20]**. In the same way as for the 2-photon force calculations, the 3-photon scattering force can be written as $\mathbf{f}_z^{(3)} = \hbar k_{ee'} \sum_{\mathbf{i},\mathbf{j}} R_{\mathbf{i},\mathbf{j}}^{(3)} \hat{\mathbf{j}}$, where, for atoms moving at velocity $\mathbf{v}$, the 3-photon scattering rate $R_{\mathbf{i},\mathbf{j}}^{(3)}$ is

$$R_{\mathbf{i},\mathbf{j}}^{(3)} = \frac{|\Omega_{ee'}|^2}{4|\tilde{\Delta}_1 - k_{ge}\hat{\mathbf{i}}\cdot\mathbf{v} - 2k_{ee'}\hat{\mathbf{j}}\cdot\mathbf{v}|^2}\frac{\Gamma}{\gamma}R_{\mathbf{i},\mathbf{j}}^{(2)}. \qquad (2)$$

We Taylor-expand $f_z^{(3)}$ near $v_z = 0$ to find the 3-photon damping coefficient $\alpha^{(3)}$. $\tilde{\Delta}_1 = \Delta_1 + i\Gamma/2$ and $\tilde{\delta}_2 = \delta_2 + i\gamma/2$; For $\Delta_1 < 0$ and $\gamma^2 = \Gamma^2 + 4\Delta_1^2$, we find $\alpha^{(3)} > 0$ for either $\delta_2 < 0$ or $\delta_2 > -\Delta_1/2$. Here the $\alpha^{(3)}$ involves only the 3-photon process and ignores the 2-photon process, light shift, and higher order process. The 3-photon cooling effect can be understood qualitatively from the |g>–|e>–|e'> Raman process. At large $|\delta_2|$, the Doppler sensitivity along the z axis becomes independent of $\delta_2$, but remains dependent on $\Delta_1$. The fact that $\alpha^{(3)}$ is positive is determined by the negative $\Delta_1$.

The discussion above has shown the cooling and trapping in type-I TC-MOT. With respect to type-II TC-MOT, $\hat{\mathbf{i}} \in \{\hat{x}, -\hat{x}, \hat{y}, -\hat{y}, \hat{z}\}$ is one of the five directions of the |g>–|e> CTBs, and $\hat{\mathbf{j}} \in \{-\hat{\mathbf{z}}\}$ is one of the $-\hat{\mathbf{z}}$ directions of the |e>–|e'> CTBs, and the analyses should be similar as the type-I TC-MOT.

In addition, compare to the cascade transitions in the $6S_{1/2}$ - $6P_{3/2}$ - $8S_{1/2}$ TC-MOT, the cascade transitions in the $6S_{1/2}$ - $6P_{3/2}$ - $7S_{1/2}$ TC-MOT is significantly simpler due to less decay channels, as shown in Fig. 1(a). It is not only profitable for the understanding of cooling and trapping mechanism (see Fig. 2(a) and the discussion) but also for the fluorescence detection. Usually, the absorption detection or laser-induced fluorescence after turning off the MOT can be used to estimate the atom number. For simplicity here we use the in-situ fluorescence measurement via a CCD camera to estimate cold cloud's size and atom numbers. In $6S_{1/2}$ - $6P_{3/2}$ - $8S_{1/2}$ TC-MOT, atoms can decay from $8S_{1/2}$ to $6S_{1/2}$ state through the cascaded $8S_{1/2}$ - $7P_{3/2}(7P_{1/2})$ - $6S_{1/2}$ and $8S_{1/2}$ - $6P_{3/2}(6P_{1/2})$ - $6S_{1/2}$ two-photon transitions, and the cascaded $8S_{1/2}$ - $7P_{3/2}(7P_{1/2})$ - $7S_{1/2}$ - $6P_{3/2}(6P_{1/2})$ - $6S_{1/2}$ and $8S_{1/2}$ - $7P_{3/2}(7P_{1/2})$ - $5D_{5/2}(5D_{3/2})$ - $6P_{3/2}(6P_{1/2})$ - $6S_{1/2}$ four-photon transitions. Even though interference filters can be employed to choose the desired wavelength, still it is a little bit difficult to estimate cold atom number from the fluorescence signal because too many branches have to be considered. Now the situation is much simpler for $6S_{1/2}$ - $6P_{3/2}$ - $7S_{1/2}$ TC-MOT, in which atoms can decay from $7S_{1/2}$ to $6S_{1/2}$ state through the cascaded $7S_{1/2}$ - $6P_{3/2}(6P_{1/2})$ - $6S_{1/2}$ two-photon transitions. The cold atom number is much more linearly dependent on the fluorescence to a certain degree. Strictly this is inaccurate, but it is a simple method to diagnose whether the TC-MOT operates and determinate the range of two-photon detuning for TC-MOT operation. For simplicity, the peak fluorescence of trapped atoms is recorded in our experiment.

## 3. Frequency offset locking system

Doppler-free spectroscopy between atomic ground and excited states, such as saturated absorption spectroscopy, polarization spectroscopy, and modulation transfer spectroscopy, provides a reference to control the frequency of 852 nm laser. The spectroscopy between excited states, such as optical-optical double resonance **[21, 22]**, double-resonance optical pumping **[23-25]**, ladder-type electromagnetically induced transparency **[26, 27]**, and two-color polarization spectroscopy **[28, 29]**, can be used to stabilize 1470 nm laser frequency. The frequency control system of cooling lasers is shown in Fig. 2. The frequency detuning $\Delta_1$ of 852 nm cooling laser is controlled by two acousto-optic modulators (AOM3 and AOM2). For example, in order to set $\Delta_1$ to be -12.5 MHz, the laser interacted with Cs vapor cell is locked to Cs $6S_{1/2}$ (F=4) - $6P_{3/2}$ (F'=4, 5) crossover line by using of polarization spectroscopic locking scheme, and the radio frequency (RF) signals applied on AOM3 and AOM2 should be 96.52 and 80 MHz, respectively. The frequency detuning $\Delta_2$ of 1470 nm cooling laser is controlled by AOM3 and a fiber-pigtailed phase-type wave-guide electro-optic modulator (EOM). For example, the RF signals applied on AOM3 and EOM are 96.52 MHz and $\Omega_{RF}$ ($\Omega_{RF}$: 90 ~ 410 MHz), respectively, and these yield that the 1470 nm laser frequency is [$\omega_2$ + ($\Omega_{RF}$ -227.5 MHz) ×852/1470], thus $\delta_2=\Delta_1+\Delta_2$ ranges from -92.2 to +93.3 MHz. Here the 1470 nm laser frequency is offset locked by the velocity-selective off-resonant two-color polarization spectroscopy [28]. The 852 nm laser detuned +227.5 MHz to $\omega_1$ pass through EOM and generate two sidebands (See Fig. 2), the -1 order sideband with frequency of ($\omega_1$+227.5 MHz-$\Omega_{RF}$) serves as pump laser for the velocity-selective off-resonance two-color polarization spectroscopy, which is used to lock 1470 nm laser. By comparison, spectroscopy arising from the +1 order sideband with frequency of ($\omega_1$+227.5 MHz+$\Omega_{RF}$) is weak because the large detuning laser only interacts with a fraction of atoms moving with large velocity which obey the Maxwell-Boltzmann distribution.

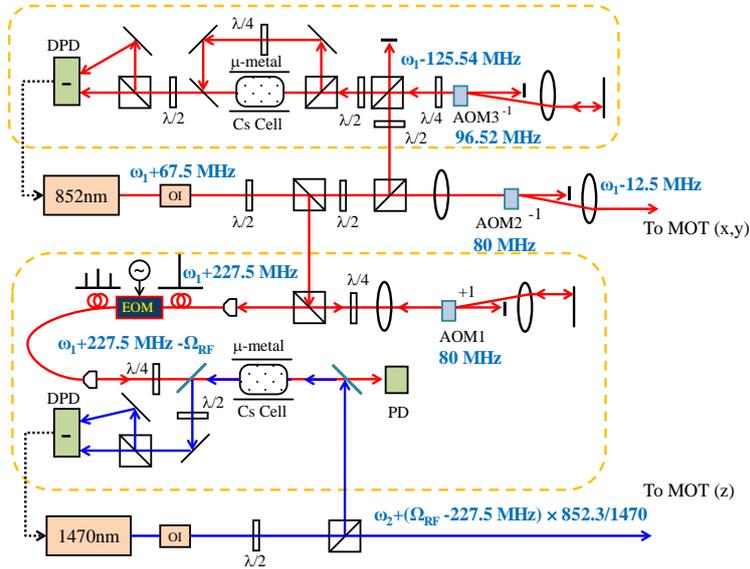

**FIG. 2.** (Color online) Schematic diagram of frequency offset locking of 852 and 1470 nm cooling lasers for the TC-MOT. Keys to figure: OI, optical isolator; PBS, polarization beam splitter cube; λ/2, half-wave plate; λ/4, quarter-wave plate; EOM, fibre-pigtailed phase-type wave-guide electro-optic modulator; PD, photodiode; DPD, differential photodiode; The frequency labelled in the figure shows the operated frequency of the lasers and the AOMs while the detuning $\Delta_1$ is set to be -12.5 MHz: $\omega_1$, Cs $6S_{1/2}$ (F=4) − $6P_{3/2}$ (F'=5) transition frequency; $\omega_2$, Cs $6P_{3/2}$ (F'=5) −$7S_{1/2}$ (F''=4) transition frequency; $\Omega_{RF}$, radio frequency applied on EOM.

Fig. 3 shows the velocity-selective off-resonance two-color polarization spectroscopy. Once the 1470 nm laser frequency is locked to the spectroscopy generated by the -1 order modulation sideband, the laser frequency should be [$\omega_2$ + ($\Omega_{RF}$ -227.5 MHz) ×852/1470]. This setup is the modification of our previous offset locking system in [10] and [28] where the RF modulation is added to 1470 nm laser. The main advantage of this locking scheme is that all frequency-shift elements are placed in the 852 nm optical path, thus to save the 1470 nm laser power. The 852 nm repumping laser which is sent along the ±y axis (not shown in Fig. 2) is locked to the $6S_{1/2}$ (F=3) - $6P_{3/2}$ (F'=4) transition by saturated absorption spectroscopic locking scheme, to avoid atoms accumulating at the $6S_{1/2}$ (F=3) ground state which does not interact with the cooling lasers.

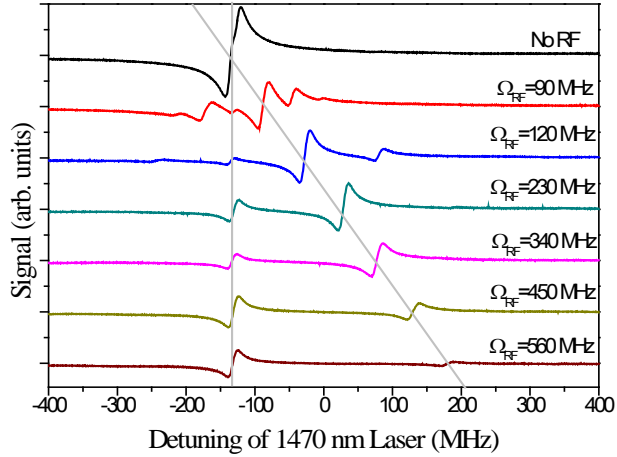

**FIG. 3.** (Color online) The velocity-selective off-resonance two-color polarization spectra versus the 1470 nm laser frequency detuning with different $\Omega_{RF}$. The 852 nm laser with frequency of ($\omega_1$+227.5 MHz) is coupled into EOM (Seen in Fig.2) to generate the modulation sidebands. The velocity-selective off-resonance two-color polarization spectroscopy can be obtained by scanning the 1470 nm laser frequency.

## 4. Experimental results and discussions

As motioned above, the fluorescence is approximately proportional to cold atom number. Here the peak fluorescence directly measured by a CCD camera in order to diagnose whether the TC-MOT operates and determine the range of two-photon detuning for the TC-MOT operation. The dependence of the peak fluorescence upon the two-photon detuning $\delta_2$, and the intensity of CTBs for type-I and type-II TC-MOTs is measured qualitatively. And some relevant discussions are also made.

*4.1. Type-I Cs TC-MOT*

Based on conventional Cs MOT, two 852 nm CTBs along the ±z directions are replaced with 1470 nm CTBs, as shown in Fig. 1(c), we named this case as the type-I Cs TC-MOT. The peak fluorescence as function of the single-photon detuning $\Delta_1$, the two-photon detuning $\delta_2$, the 852 and 1470 nm CTBs' power, and the repumping beams' power are shown in Fig. 4.

The significant characteristic of the TC-MOT is that it can cool and trap atoms on both the red- and blue-detuning sides of the two-photon resonance. As motioned above, the two-photon detuning $\delta_2$ is controlled (from -90 to +90 MHz) by EOM. Two typical false-color fluorescence images of cold cloud are shown as the insets of Fig. 4(d), corresponding to the data points for $\delta_2$ = +28.4 and -36.6 MHz with a maximum repumping power, respectively. The sizes of the two clouds are about 0.8 mm (z) × 0.3 mm (x, y) and 0.9 mm (z) × 0.2 mm (x, y), respectively. Typical atom number is estimated to be ~ 5×10$^6$, and the corresponding atomic density is 6.9×10$^{10}$~1.3×10$^{11}$/cm$^3$.

Fig. 4(a) shows the peak fluorescence of atoms trapped in the type-I Cs TC-MOT as a function of $\delta_2$ with different 1470 nm CTBs' power, while the 852 nm CTBs' total power is 4×6.10 mW, and the single-photon detuning is $\Delta_1$ = -12.5 MHz. On the red-detuning side of the two-photon resonance, as the 1470 nm CTBs' power increases, the range of $\delta_2$ for TC-MOT operation broadens and is red-shifted. On the blue-detuning side, TC-MOT works with $\delta_2$ > 12 MHz, and the range broadens when the 1470 nm CTBs' power increases. The required CTBs' power for TC-MOT operation on the blue-detuning side is less than that on the red-detuning side. One point should be addressed here, this is inversed to that in 852+795 nm Cs TC-MOT in **[9]**, in which the required CTBs' power for TC-MOT operation on the blue-detuning side is larger. The different behaviors between 852+1470 nm and 852+795 nm TC-MOTs are probably due to following reasons: (1) The photon momentum of 1470 nm is less than that of 795 nm, thus the scattering force in former TC-MOT is weaker than that in latter one, so it is more difficult to cool and trap atoms in $\delta_2 < 0$ region. In other words, more optical power is needed at the blue-detuning side for TC-MOT operation; (2) The decay channels in former TC-MOT are much less than that in latter, in detail, the decay branching ratio for $7S_{1/2}$ - $6P_{3/2}$ channel is ~ 65% and that for $8S_{1/2}$ - $6P_{3/2}$ channel is ~ 37%, hence the Raman process in former TC-MOT is more pure and the 3-photon scattering rate is higher in $\delta_2 > 12$ MHz region; (3) In Eq. (2), the 3-photon scattering rate is inversely proportional to $\left| \tilde{\Delta}_1 - k_{ge} \hat{\mathbf{i}} \cdot \mathbf{v} - 2 k_{ee'} \hat{\mathbf{j}} \cdot \mathbf{v} \right|^2$, consequently, it is large due to small $k_{ee'}$ along the z direction. In addition, the former velocity capture range is larger than that of the latter.

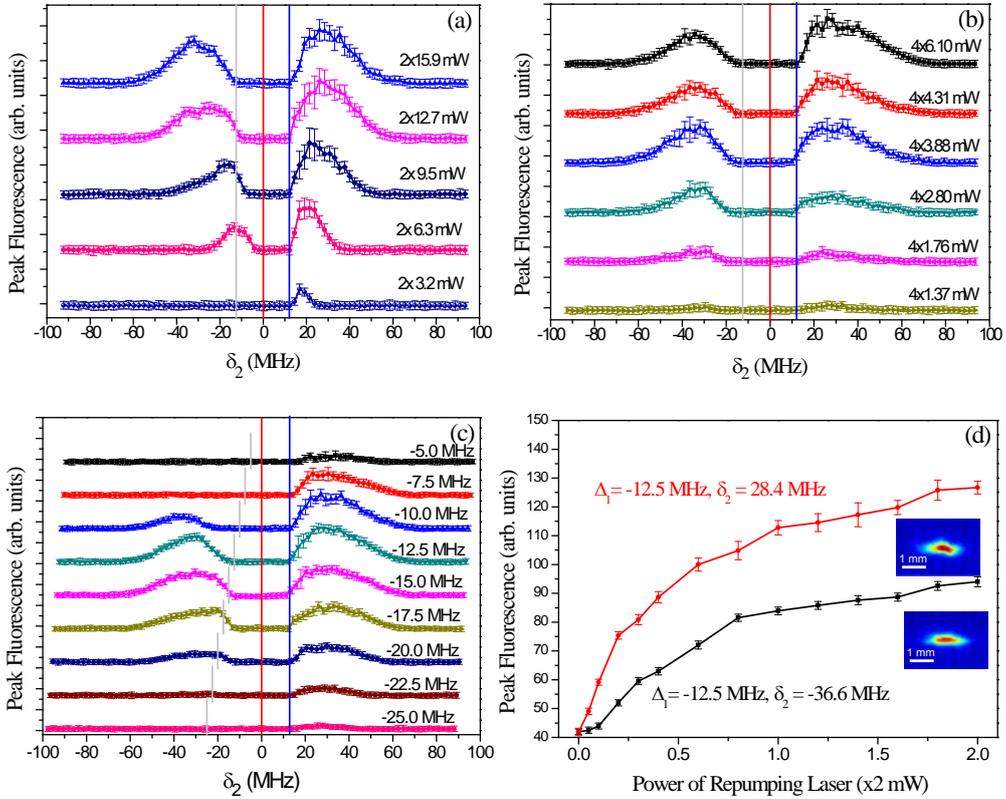

**FIG. 4.** (Color online) The peak fluorescence of Cs atoms trapped in the type-I Cs TC-MOT as a function of the two-photon detuning $\delta_2$ (a) with different 1470 nm CTBs' power, while the 852 nm CTBs' total power is 4×6.10 mW, and the single-photon detuning is $\Delta_1$ = -12.5 MHz; (b) with different 852 nm CTBs' power, while the 1470 nm CTBs' total power is 2×20.0 mW, and the single-photon detuning is $\Delta_1$ = -12.5 MHz; (c) with different single-photon detuning $\Delta_1$, while the 1470 nm CTBs' total power is 2×20.0 mW, and the 852 nm CTBs' total power is 4×6.10 mW. The repumping beams' total power is 2×2.0 mW for (a), (b) and (c). The vertical grey line represents $\Delta_2$ = 0, the vertical red line represents $\delta_2$ = 0, and the vertical blue line represents $\delta_2 \approx +12$ MHz. (d) The peak fluorescence of Cs atoms trapped in type-I TC-MOT as a function of the 852 nm repumping beams' power. The CTBs' powers are 4×6.10 mW for 852 nm and 2×20.0 mW for 1470 nm, respectively. The insets are fluorescence images of cold cloud for the two different $\delta_2$ with a maximum repumping beams' power (2×2.0 mW).

Fig. 4(b) shows the peak fluorescence of atoms trapped in type-I Cs TC-MOT as a function of $\delta_2$ with different 852 nm CTBs' power, while the 1470 nm CTBs' total power is 2×20.0 mW, and the single-photon detuning is $\Delta_1$ = -12.5 MHz. As the 852 nm CTBs' power increases, the range of $\delta_2$ for TC-MOT operation on both the red- and blue-detuning sides does not change much (neither shift nor broaden much) due to force balance along the z direction. The 852 nm CTBs are orthogonal to the 1470 nm CTBs, as a result, when the 852 nm CTBs' power increase, the atom number increase, but the force balance along the z direction does not break.

Fig. 4(c) shows the peak fluorescence of atoms trapped in type-I Cs TC-MOT as a function of $\delta_2$ for atoms in type-I TC-MOT with different single-photon detuning $\Delta_1$, while the 1470 nm CTBs' total power is 2×20.0 mW, and the 852 nm CTBs' total power is 4×6.10 mW. As the change of single-photon detuning $\Delta_1$, the peak fluorescence has an optimized intensity at $\Delta_1$ = -12.5 MHz. Larger or smaller than this value, the peak fluorescence reduces. Note that the single-photon detuning $\Delta_1$ seems not to shift the range of $\delta_2$ for TC-MOT operation due to perpendicular between 852 and 1470 nm CTBs. The gray lines representing $\Delta_2$ =0 ($\delta_2=\Delta_1$) in each curve provide a most direct impression why we consider the two-photon detuning $\delta_2$ instead of the detuning $\Delta_2$.

Fig. 4(d) shows the peak fluorescence of atoms trapped in type-I Cs TC-MOT as a function of the 852 nm repumping beams' power. The single-photon detuning is $\Delta_1$ = -12.5 MHz. The two photon detunings are $\delta_2$ = 28.4 MHz for upper curve, and $\delta_2$ = -36.6 MHz for lower curve. The CTBs' powers are 4×6.10 mW for 852 nm and 2×20.0 mW for 1470 nm, respectively. The insets are fluorescence images of cold cloud for the two different $\delta_2$ with a maximum repumping beams' power (2×2.0 mW). As the repumping beams' power increases, the peak fluorescence increases and tends to be saturated on both red- and blue-detuning sides.

*4.2. Type-II Cs TC-MOT*

Based on the type-I Cs TC-MOT, one of the 1470 nm CTBs along the ±z directions is replaced with one 852 nm CTB, as shown in Fig. 1(d), we named this case as the type-II Cs TC-MOT.

Fig. 5 shows the peak fluorescence of atoms trapped in type-II TC-MOT as a function of $\delta_2$ with different 852 nm CTB's power along the +z direction, while the 1470 nm CTB's power is 9.5 mW. The 852 nm CTBs' total power in the x-y plane is 4×1.99 mW (a) and 4×6.10 mW (b). The single-photon detuning is $\Delta_1$ = -12.5 MHz. The type-II TC-MOT can also cool and trap atoms on both the red- and blue-detuning sides of the two-photon resonance. As the 852 nm CTB's power along the +z direction increases, on the red-detuning side, the range of for TC-MOT operation shifts close to resonance until the power is larger than a threshold value 2.66 mW in Fig. 5(a) and 5.80 mW in Fig. 5(b), respectively. On the blue-detuning side, the TC-MOT works with $\delta_2$ > 8 MHz, the range of $\delta_2$ for TC-MOT operation narrows, and the fluorescence disappears when the 852 nm CTB's power is larger than a threshold value 1.30 mW in Fig. 5(a) and 4.09 mW in Fig. 5(b), respectively.

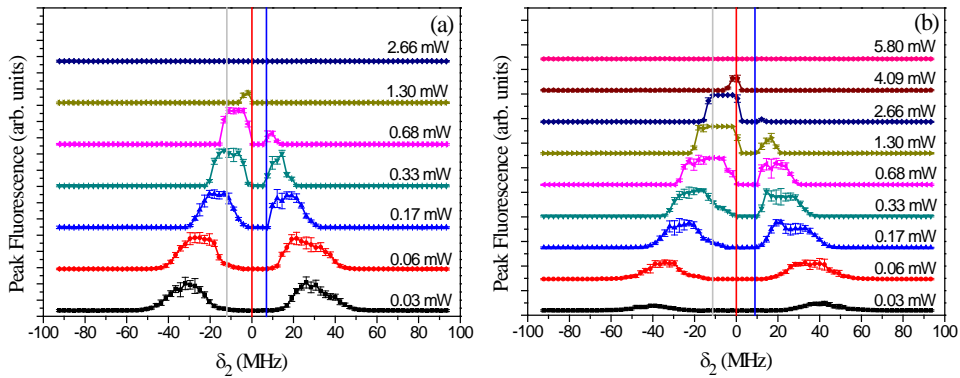

**FIG. 5.** (Color online) The peak fluorescence of Cs atoms trapped in the type-II Cs TC-MOT as a function of the two-photon detuning $\delta_2$ with different 852 nm CTB power along the +z direction, while the 1470 nm CTB power is 9.5 mW. The 852 nm CTBs' total power in the x-y plane is 4×1.99 mW (a) and 4×6.10 mW (b). The single-photon detuning is $\Delta_1$= -12.5 MHz. The vertical grey line represents $\Delta_2 = 0$, the vertical red line represents $\delta_2 = 0$, and the blue vertical line represents $\delta_2 \approx +8$ MHz.

There are two distinguished characteristics: (1) In each of the figures, as the 852 nm CTB's power along the +z direction increases, the range of $\delta_2$ for TC-MOT operation noticeably shifts. This is because the scattering force arising from 852 and 1470 nm CTBs should be equal along the z direction. For the 852 nm CTB with $\Delta_1$ =-12.5 MHz, as its power increases, the scattering force increases. For the 1470 nm CTB with fixed power, as 852 nm CTB's scattering force increases, the detuning $\delta_2$ has to shift close to resonance to keep the balance between the scattering forces. (2) For TC-MOT operation, the 852 nm CTB's upper limit power along the +z direction in Fig. 5 (b) is higher than the power in Fig. 5 (a). One reason is that the scattering force of 1470 nm CTB is higher in Fig. 5 (b) due to larger Rabi frequency $\Omega_{ge}$ according to Eq. 1 and 2 for higher 852 nm CTBs' power in the x-y plane. Hence the 852 nm CTB's upper limit power along the +z direction in Fig. 5 (b) should be higher to keep the force balance. Another possible reason is the misalignment of the two beams (<5 mrad) along the z axis, producing a resultant force in the x-y plane, which push the atoms away from the center of TC-MOT. Although the scattering forces between 852 nm and 1470 nm CTBs keep balance along the z axis, to keep the force balance in the x-y plane, the binding force from 852 nm CTBs in the x-y plane should be large enough to against this resultant force. In other words, the 852 nm CTB's upper limit power along the +z direction in Fig. 5 (b) is higher due to the stronger binding force in the x-y plane.

Fig. 6 shows the peak fluorescence of Cs atoms trapped in the type-II Cs TC-MOT as a function of $\delta_2$ with other different parameters. In Fig. 6 (a), with higher 1470 nm CTB's power, to keep the force balance between scattering forces of 852 and 1470 nm CTB, the two-photon detuning $\delta_2$ for TC-MOT operation has to shift away from resonance. In Fig. 6 (b), with higher 852 nm CTBs' power in the x-y plane, the scattering force of 1470 nm CTB is higher due to larger Rabi frequency $\Omega_{ge}$ according to Eq. 1 and 2. So the two-photon detuning $\delta_2$ for TC-MOT operation has to shift away from resonance to keep the force balance. Moreover, this feature is different from that in type-I TC-MOT illustrated in Fig. 5 (b). In that case, the range of $\delta_2$ for TC-MOT operation on both the red- and blue-detuning sides does not change much due to force balance along the z direction. In Fig. 6 (c), larger single-photon detuning $\Delta_1$ will weaken the scattering force arising from 852 nm CTBs both in the x-y plane and along the z axis, hence the $\delta_2$ for TC-MOT operation should slightly shift away from resonance to keep the scattering force balance. In sum, one needs to carefully adjust all the parameters, such as the 1470 nm CTB's power, the 852 nm CTB's power both in the x-y plane and along the z direction, the single-photon detuning $\Delta_1$, and the two-photon detuning $\delta_2$, to keep the scattering force balance and trap more atoms.

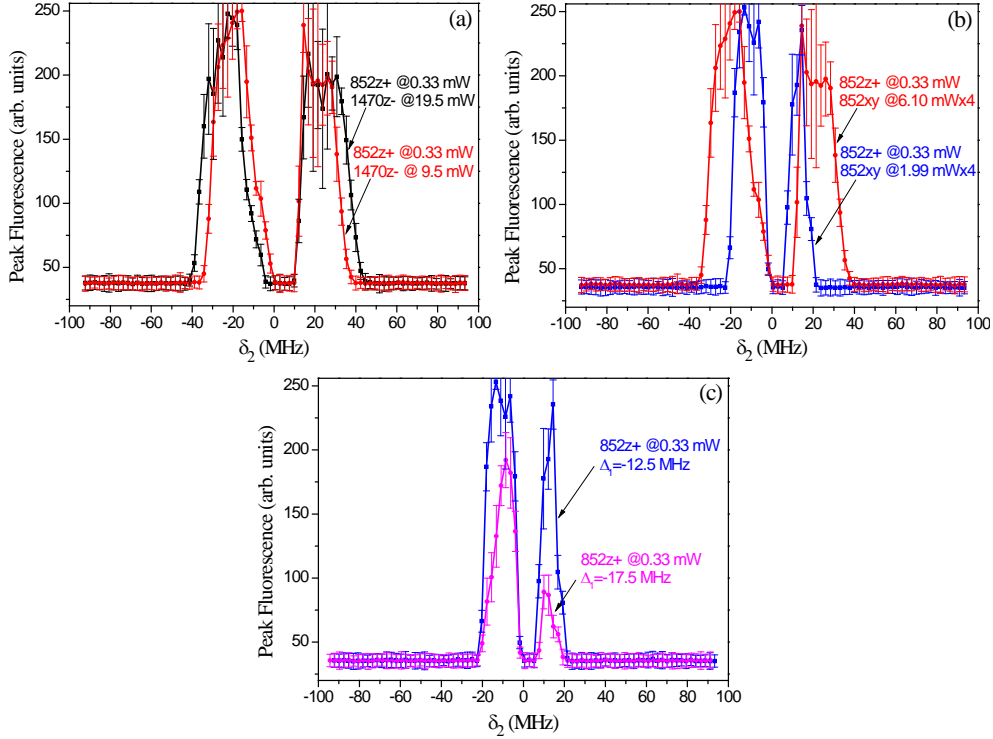

**FIG. 6.** (Color online) The peak fluorescence of Cs atoms trapped in the type-II Cs TC-MOT as a function of the two-photon detuning $\delta_2$ with other different parameters. (a) With different 1470 nm CTB; (b) With different 852 nm CTBs' power in the x-y plane; (c) With different single-photon detuning $\Delta_1$. The details are: (i) black line, $\Delta_1$ = -12.5 MHz, $P_{CTB\ xy}$=4×6.10 mW, $P_{CTB\ z+}$=0.33 mW, $P_{CTB\ z-}$=19.5 mW; (ii) red line, $\Delta_1$ = -12.5 MHz, $P_{CTBxy}$=4×6.10 mW, $P_{CTB\ z+}$=0.33 mW, $P_{CTB\ z-}$=9.5 mW; (iii) blue line, $\Delta_1$ = -12.5 MHz, $P_{CTB\ xy}$=4×1.99 mW, $P_{CTB\ z+}$=0.33 mW, $P_{CTB\ z-}$=9.5 mW; (ix) magenta line, $\Delta_1$ = -17.5 MHz, $P_{CTB\ xy}$=4×1.99 mW, $P_{CTB\ z+}$=0.33 mW, $P_{CTB\ z-}$=9.5 mW.

## 5. Conclusion

In conclusion, a novel Cs TC-MOT, in which the optical forces due to photon scattering of the Cs $6P_{3/2}$ (F'=5) - $7S_{1/2}$ (F"=4) excited-state transition in the Cs $6S_{1/2}$ - $6P_{3/2}$ - $7S_{1/2}$ (852 + 1470 nm) ladder-type system are partially employed, has been proposed and experimentally investigated. One of the three pairs of 852 nm CTBs in a conventional Cs MOT is replaced with a pair of the 1470 nm CTBs (type-I) or with one 852 nm beam plus another counter-propagating 1470 nm beam (type-II). Both the type-I and type-II Cs TC-MOTs can cool and trap atoms on both the red-detuning and blue-detuning sides of the two-photon resonance. We measured and analyzed qualitatively the dependence of peak fluorescence on the two-photon detuning, the intensity of CTBs in type-I and type-II TC-MOT.

These results not only provided optimized experimental parameters to trap atoms, but also provided helpful evidence to deeply investigate the mechanism of cooling and trapping atoms in TC-MOT. The experiment demonstrated in this work may have wide applications, such as the background-free detection of trapped atoms, the laser cooling of Rydberg atoms in a cascade atomic configuration, and the assisting laser cooling of certain atomic species which require cooling lasers at inconvenient wavelengths. Especially, type-II TC-MOT is an alternative scheme to cool and trap atoms and pump the atomic ensemble for the practical application in the photon-pair sources compatible with the ensemble-based quantum memory and the long-distance quantum communication via optical fiber. Note that high optical depth is profitable to this kind of photon-pair generation based on four-wave mixing. Higher power and large spot size can effectively enlarge the optical depth. In our experiment, the 1470 nm beam size is smaller than 852 nm beam size in order to enhance the optical intensity, but this may reduce the velocity capture range. Moreover, the technique which used to enlarge optical depth in 2D MOT [30], like a cylindrical quadrupole field, large atom number, a temporally dark and compressed MOT, and Zeeman-state optical pumping, will also be used to achieve a high optical depth.

## Acknowledgments

This work is supported by the National Major Scientific Research Program of China (Grant No. 2012CB921601), the National Natural Science Foundation of China (Grant Nos. 61475091, 11274213, 61205215, and 61227902), and Research Program for Sci and Tech Star of Tai Yuan, Shan Xi, China (Grant No.12024707).

## References

[1] Raab E L, Prentiss M, Cable A, Chu S, and Pritchard D E 1987 *Phys. Rev. Lett.* **59**, 2631-4
[2] Phillips W D 1998 *Rev. Mod. Phys.* **70**, 721-41
[3] Chu S, Hollberg L, Bjorkholm J E, Cable A, and Ashkin A 1985 *Phys. Rev. Lett.* **55**, 48-51
[4] Dalibard J and Cohen-Tannoudji C1989 *J. Opt. Soc. Am. B* **6**, 2023-45
[5] Hack J, Liu L, Olshanii M, and Metcalf H 2000 *Phys. Rev. A* **62**, 013405
[6] Curtis E A, Oates C W, and Hollberg L 2001 *Phys. Rev. A* **64**, 031403(R)
[7] Malossi N, Damkjær S, Hansen P L, Jacobsen L B, Kindt L, Sauge S, and Thomsen J W 2005 *Phys. Rev. A* **72**, 051403(R)
[8] Mehlstäubler T E, Moldenhauer K, Riedmann M, Rehbein N, Friebe J, Rasel E M, and Ertmer W 2008 *Phys. Rev. A* **77**, 021402(R)
[9] Wu S, Plisson T, Brown R C, Phillips W D, and Porto J V 2009 *Phys. Rev. Lett.* **103**, 173003
[10] Yang B D, Liang Q B, He J, and Wang J M 2012 *Opt. Express* **20**, 11944-52
[11] Wu S, Brown R C, Phillips W D, and Porto J V 2011 *Phys. Rev. Lett.* **106**, 213001
[12] Sangouard N, Simon C, Riedmatten H De, and Gisin N 2011 *Rev. Mod. Phys.* **83**, 33-80
[13] Chanelière T, Matsukevich D N, Jenkins S D, Kennedy T A B, Chapman M S, and Kuzmich A 2006 *Phys. Rev. Lett.* **96**, 093604
[14] Burnham D C and Weinberg D L 1970 *Phys. Rev. Lett.* **25**, 84-87
[15] Jia X J, Yan Z H, Duan Z Y, Su X L, Wang H, Xie C D, Peng K C 2012 *Phys. Rev. Lett.* **109**, 253604
[16] Balić V, Braje D A, Kolchin P, Yin G Y, and Harris S E 2005 *Phys. Rev. Lett.* **94**, 183601
[17] Kolchin P, Du S, Belthangady C, Yin G Y, and Harris S E 2006 *Phys. Rev. Lett.* **97**, 113602
[18] Du S, Kolchin P, Belthangady C, Yin G Y, and Harris S E 2008 *Phys. Rev. Lett.* **100**, 183603
[19] Theodosiou C E 1984 *Phys. Rev. A* **30**, 2881-909
[20] Tollett J J, Chen J, Story J G, Ritchie N W M, Bradley C C, and Hulet R G 1990 *Phys. Rev. Lett.* **65**, 559-62
[21] Sasada H 1992 IEEE *Photon.Technol. Lett.* **4**, 1307-9
[22] Boucher R, Breton M, Cyr N, and Têtu M 1992 *IEEE Photon. Technol. Lett.* **4**, 327-9
[23] Moon H S, Lee W K, Lee L, and Kim J B 2004 *Appl. Phys. Lett.* **85**, 3965-7

[24] Yang B D, Zhao J Y, Zhang T C, and Wang J M 2009 *J. Phys. D: Appl. Phys.* **42**, 085111
[25] Wang J, Liu H F, Yang G, Yang B D, and Wang J M 2014 *Phys. Rev. A* **90**, 052505
[26] Gea-Banacloche J, Li Y, Jin S, and Xiao M 1995 Phys. Rev. A **51**, 576-84
[27] Wang J, Liu H F, Yang B D, He J, and Wang J M 2014 Meas. Sci. Technol. **25**, 035501
[28] Yang B D, Wang J, Liu H F, He J, and Wang J M 2014 *Opt. Commun.* **319**, 174-7
[29] Liu H F, Wang J, Yang G, Yang B D, He J, and Wang J M 2014 *Chinese J. Lasers* **41**, 0715004 (in Chinese)
[30] Hsiao Y F, Chen H S, Tsai P J, and Chen Y C 2014 *Phys. Rev. A* **90**, 055401